\documentclass[aps, singlecolumn, showpacs]{revtex4-2}
\usepackage{amsmath}
\begin{document}
\title{ Symmetry of Dirac two-oscillator system, gauge-invariance, and Landau problem}
\author{S C Tiwari}
\affiliation{Department of Physics, Institute of Science, Banaras Hindu University, Varanasi 221005, and \\ Institute of Natural Philosophy \\
Varanasi India\\ Email address: $vns\_sctiwari@yahoo.com$ \\}
\begin{abstract}
 Role of gauge symmetry in the proton spin problem has intricate and unresolved aspects. One of the interesting approaches to gain physical insights is to explore tha Landau problem in this context. A detailed study using the group theoretic method to understand the Landau problem establishes the significance of the gauge transformation intimately related with the space translation symmetry. An important implication of this result is that the E(2)-like Wigner's little group for massless particles could throw more light on the question of gauge symmetry in QED and QCD. A generalized Landau-Zeeman Hamiltonian is proposed in which Dirac two-oscillator system and the symmetry of the group SO(3,2) become important. It is argued that nontrivial topology of pure gauge field holds promise to resolve the unsettled questions.
\end{abstract}

\maketitle
 
\section{Introduction}
The debate on the proton spin decomposition controversy brought into sharp focus the issue of gauge symmetry: choice of a gauge, gauge-fixing, gauge-invariance, and meaning of gauge-covariant operators in quantum field theory \cite{1,2,3,4,5,6}. It seems the importance of the Landau problem in connection with the proton spin was first pointed out by Wakamatsu, see for example,  \cite{7}. In a recent paper \cite{8}  based on their study on the Landau problem the authors conclude: "Undoubtedly, the present analysis provides us with clear and concrete example in which the gauge symmetry is just a redundancy of the description with little physical substance." This conclusion merits critical assessment. First, we point out that pure gauge field could be attributed physical significance based on the topological considerations \cite{9,10}. Secondly, as reminded in \cite{2} the covariant quantization and gauge-fixing are nontrivial issues in quantum field theories (QED and QCD); Leader \cite{11} rightly noted the necessity of an indefinite-metric space, physical states and physical observables. Recently, the role of the  states of negative norm in the definition of the physical states in \cite{11} has been re-examined \cite{12,13}. 

Though the nonrelativistic time-independent Landau problem has limited implications on the proton spin problem in QCD, it does offer new physical insights. The role of Landau problem could be approached in a different way than that of \cite{7,8} based on symmetry considerations and the Zeeman effect for harmonic oscillator \cite{14}. We term this problem Landau-Zeeman problem. Dirac's remarkable respresentation of SO(3,2) group \cite{15} and two-oscillator system have been discussed recently \cite{16} gaining new insights on the two-oscillator system of Dirac. A thorough study on the symmetries of the Landau problem contrasting the problem with the two dimensional (2D) isotropic harmonic oscillator (IHO) and Dirac oscillators shows the central extension of the Euclidean group in 2D denoted by $\bar{E}(2)$  as a dynamical symmetry group and symplectic group Sp(2R) as spectrum generating group for the Landau problem irrespective of the choice of the gauge. We have used some of the interesting ideas in \cite{17,18,19}, the method of group contraction \cite{20}, and the maximal kinetic group for 1D oscillator \cite{21}. An attempt is made to put the work of Wakamatsu and Hayashi \cite{8} in this perspective. We find that gauge symmetry plays important role in connection with the translational symmetry, and gauge symmetry/transformation is not redundant.

However, SO(3,2) group is not relevant as such for the Landau problem. We propose a new Landau-Zeeman system where Dirac's SO(3,2) symmetry group becomes natural. The important contributions of Kim and his group \cite{17} and Kibler \cite{18} on the symmetry group for two-oscillator system prove useful in this connection.  Dirac's remarkable representation of the group SO(3,2) is shown to be a symmetry group for this system using the equivalence of this group with Sp(4,R) proved by Dirac \cite{15}. A natural generalization in the context of E(2)-like Wigner's little group for massless particles, and the suggested relevance of SO(3,2) group in QED and QCD seems to offer a new approach to the spin decomposition controversy.

In the next section, the issue of the gauge symmetry in the context of the proton spin problem is briefly reviewed.  In section 3, the group theoretic approach on the Landau problem is presented, and the conclusions arrived at in \cite{8} are discussed in this new framework \cite{16}.  In section 4, in order to gain new insights on the Landau problem and gauge symmetry a generalized Landau-Zeeman Hamiltonian is proposed. In section 5 wider symmetry group SO(3,2) and Dirac two-oscillator system are discussed. Concluding remarks constitute the last section.

\section{\bf Covariant quantization and gauge symmetry}

The quantization of the fields and local gauge field theories are extensively discussed in many textbooks on quantum field theory. Proton spin puzzle and nucleon spin decomposition controversy \cite{4,5} brought into sharp focus certain conceptual issues related with the local gauge symmetry. A brief review is presented in this section.

The classical theory of electromagnetism (CED) is founded on the experimental laws abstracting the physical concept of the electromagnetic fields $F_{\mu\nu}$ from the force laws. Relativistic invariance (covariance) of the Maxwell field equations predates Einstein's special theory of relativity. Introduction of the electromagnetic potentials $A_\mu$ as such would appear superfluous. However, variational formulation using the action integral crucially depends on $A_\mu$ treated as independent dynamical field variable in the action principle. In addition to the space-time Lorentz symmetry there would arise a gauge symmetry if $A^\mu$ is introduced. A gauge transformation
\begin{equation}
A_\mu ~ \rightarrow ~ A_\mu +\partial_\mu \chi
\end{equation}
leaves the EM field tensor and the action
\begin{equation}
S = -\frac{1}{4}\int F^{\mu\nu} F_{\mu\nu} ~d^4 x
\end{equation}
invariant. Invariance of (2) under the inhomogeneous Lorentz transformations leads to the conservation laws \cite{22}. Consider the canonical energy-momentum tensor $T^{\mu\nu}$ that turns out to be neither symmetric nor gauge invariant in contrast to the standard expressions for energy and momentum densities that depend only on the fields ${\bf E}, {\bf B}$ being gauge invariant. Making a mathematical transformation
\begin{equation}
E^{\mu\nu} = T^{\mu\nu} + \partial_\lambda (F^{\mu\lambda} A^\nu)
\end{equation}
the new quantity $E^{\mu\nu}$ is symmetric and $E^{00}, E^{ij}$ agree with the standard energy and momentum densities respectively of the EM fields. Note that the total energy and momentum calculated from the integration of $T^{\mu\nu}$ over the whole space agree with the standard expressions as the surface integrals of the divergence terms vanish. Interplay of the gauge transformations could also be seen if we consider the Maxwell field equation 
\begin{equation}
\partial_\mu F^{\mu\nu} =0
\end{equation}
that gives the vector wave equation for $A^\mu$ using (1) and assuming the Lorentz gauge condition
\begin{equation}
\partial_\mu A^\mu =0
\end{equation}
Eq.(5) is manifestly Lorentz covariant and gauge invariant if 
\begin{equation}
\partial^\mu \partial_\mu \chi =0
\end{equation}.
Choice of the gauge (5) restricts allowed gauge transformations. Action (2) is gauge invariant, however, using a Lagrange multiplier one can fix the gauge (5) in the action itself. If we make the gauge symmetry to be an essential ingredient in CED then the definition of $A^\mu$ through 
\begin{equation}
F^{\mu\nu} =\partial^\mu A^\nu - \partial^\nu A^\mu
\end{equation}
shows that $A^\mu$ under the Lorentz transformations transforms as a vector only upto a gauge transformation \cite{23}. This point is re-emphasized \cite{4} in connection with the proton spin problem. Almost everything described here is a textbook matter; the reason to summarize it here is that most of the gauge symmetry related issues in QED have CED background \cite{2}.

In QED treated as a gauge field theory the question of gauge fixing is important in both canonical quantization and functional integral quantization. In the functional integral quantization, Fadeev-Popov trick has proved useful for gauge fixing. However, the gauge invariant S-matrix has to be constructed using the projection on the sub-space that obeys the transversality. 

In principle, manifestly Lorentz covariant gauge (5) should be preferred. Unfortunately, there are technical impediments in the quantization of a massless vector field. Weinberg \cite{23}, see sections 8.2 and 8.3, lists various gauges and chooses the Coulomb gauge that respects manifest rotational invariance
\begin{equation}
{\bf \nabla}.{\bf A} =0
\end{equation}
In connection with the angular momentum, the import of the gauge fixing in the gauge invariant covariant quantization is nicely discussed in \cite{12}. The review \cite{4} notes that gauge fixing fields result into an infinity of gauge fixed Lagrangians, and the determination of physical quantities/observables is not easy. Authors \cite{4} define gauge invariance as the property of a quantity invariant under a c-number gauge transformation, and gauge independence as a property of a quantum variable whose value is independent of the method of the gauge fixing. Gauge invariant extension in connection with proton spin decomposition based on \cite{1} is also suggested \cite{4}. In the framework of the Landau problem the notion of gauge invariant (covariant) extension is analyzed in \cite{8}. Insightful remarks on gauge invariant extension in QCD could be found in \cite{7,8}. However, it may be pointed out that compared to QED a non-abelian gauge field theory (QCD) has additional issues, for example, the ghost fields. Therefore, nonrelativistic quantum mechanical Landau problem seems to have only a superficial analogy to gauge invariant extension in QED and QCD. Group theoretic approach and symmetry considerations put forward in the present paper on the Landau problem may prove more useful as we argue in section 5.

\section{\bf Symmetry group and gauge independence: Landau problem}

Let us consider the Hamiltonian for the Landau problem in the symmetric gauge 
\begin{equation}
H_L = \frac{1}{2m} ({\bf p} +\frac{e{\bf A}}{c})^2 = H_o +\omega_L L
\end{equation}
where 
\begin{equation}
H_o = \frac{1}{2m} (p_x^2 +p_y^2) +\frac{1}{2} m \omega_L^2(x^2 + y^2)
\end{equation}
It may be noted that for a uniform magnetic field along z-axis one has the freedom to choose the vector potential either in the symmetric gauge 
\begin{equation}
{\bf A} =\frac{B}{2} (-y \hat{x} +x \hat{y} +0\hat{z})
\end{equation}
or in the Landau gauge
\begin{equation}
{\bf A} = B (0\hat{x} +x\hat{y} +0\hat{z} )
\end{equation}
For the considerations of the symmetry group the salient features of the Landau problem are as follows. Though 2D IHO Hamiltonian (10) present in the Landau Hamiltonian $H_L$ implies some role of 2D oscillator the crucial difference due to the angular momentum term results into a nontrivial aspect related with the infinite degeneracy: the dynamical symmetry group SU(2) for 2D IHO \cite{19}, as it is, cannot occur in the Landau problem. The choice of the gauge introduces additional symmetry: the gauge transformation (1) in the symmetric gauge (11) leads to the two Landau gauges. The remarkable thing is that in all the three cases the Landau problem has energy eigenvalues corresponding to 1D HO with angular frequency $\omega_c =\frac{e B}{m c}=2 \omega_L$.

An interesting approach to study the symmetry of the Landau problem suggested by Dulock and McIntosh \cite{14} is to introduce additional attractive harmonic potential with angular frequency $\omega$ in $H_L$, i. e. Zeeman effect
\begin{equation}
H_Z = H_L +\frac{1}{2} m \omega^2 (x^2 + y^2)
\end {equation}
Considering the monomials $(x, y, p_x, p_y)$ the eigenvalues and eigenvectors are calculated. The eigenvectors are
\begin{equation}
u = {[(\omega_L^2+\omega^2) m]}^\frac{1}{2} (x+iy) +i m^{-\frac{1}{2}} (p_x +i p_y)
\end{equation}
\begin{equation}
v =  {[(\omega_L^2+\omega^2) m]}^\frac{1}{2} (x-iy) +i m^{-\frac{1}{2}} (p_x -i p_y)
\end{equation}
The eigenvalues are obtained in terms of
\begin{equation}
\lambda_1 = (\omega^2 + \omega_L^2)^\frac{1}{2} + \omega_L
\end{equation}
\begin{equation}
\lambda_2 = (\omega^2 + \omega_L^2)^\frac{1}{2} - \omega_L
\end{equation}
Let us define R by $R\lambda_1 =\lambda_2$. Linear combinations of the products of the quantities $(u, v)$ constitute the constants of motion 
\begin{equation}
K=\frac{(u^R v^* + u^{*R} v)}{\sqrt{R} {(u u^*)}^{\frac{(R-1)}{2}}}
\end{equation}
\begin{equation}
L=i \frac{(u^R v^* - u^{*R} v)}{\sqrt{R} {(u u^*)}^{\frac{(R-1)}{2}}}
\end{equation}
\begin{equation}
D=\frac{(u u^* - R v v^*)}{R}
\end{equation}
These constants of motion together with the Hamiltonian $H_Z$ satisfy SU(2) Lie algebra.

To understand the symmetries of the Landau problem we adopted a novel approach \cite{16} based on the group contraction \cite{20}. In
this method one deals with the Lie algebra of the group: if the basis elements of the linear space defining the Lie algebra change by a non-singular transformation then the transformed Lie algebra describes the isomorphic algebra to the original one. If a singular transformation is made and the transformed generators fulfil the conditions of a Lie algebra one has a new group. We utilize an important property of the contraction procedure: the transformation of basis elements is made in a sequence depending on some parameters $\epsilon_i \rightarrow 0$, the contraction may lead to different groups depending on the sequence. A transformation is singular if all the generators depending on $\epsilon$ tend to zero either in the same way as $\epsilon \rightarrow 0$ or some of them vanish faster than $\epsilon \rightarrow 0$. The interesting aspect of the group SU(2) relevant here is as follows. The Lie algebra of SU(2) group with the generators $F_i,~i=1,2,3$ is known to be
\begin{equation}
[F_1, F_2] =F_3, ~ [F_2, F_3] = F_1, ~ [F_3, F_1] = F_2
\end{equation}
If $F_i \rightarrow F_i^c = \epsilon F_i, i=1,2$ and $F_3 \rightarrow F_3^c = F_3$ then the algebra of the contraction group is that of E(2) letting the limit $\epsilon \rightarrow 0$
\begin{equation}
[F_1^c, F_2^c] =0, ~ [F_2^c, F_3^c] = F_1^c, ~ [F_3^c, F_1^c] = F_2^c
\end{equation}
For the second kind of sequence $F_i \rightarrow F_i^c = \epsilon F_i, i=1,2$ and $ F_3 \rightarrow F_3^c= \epsilon^2 F_3$, in the limit $\epsilon \rightarrow 0$ we have the following contraction algebra
\begin{equation}
[F_1^c, F_2^c] = F_3^c, ~ [F_2^c, F_3^c] = [F_3^c, F_1^c]=0
\end{equation}
Note that the Heisenberg commutators $[x_i, p_j] = i\hbar \delta_{ij}$ define the Heisenberg algebra; the contraction algebra (23) is the Heisenberg algebra.

The limiting cases for SU(2) in $H_Z$ are treated using the method of group contraction; for details we refer to \cite{16}. The limiting case $\omega_L \rightarrow 0$ corresponds to a trivial case of a nonsingular transformation leading to $SU(2) \rightarrow SU(2)$ symmetry group. The second limiting case $\omega \rightarrow 0$ represents a singular transformation in which $R \rightarrow 0$. In the contraction method we let $\epsilon = R^\frac{1}{2}$ and taking the limit $\epsilon \rightarrow 0$ on the transformed generators $K \rightarrow \epsilon K, ~L \rightarrow \epsilon L,~ D \rightarrow \epsilon^2 D$ of SU(2) we arrive at the Heisenberg algebra \cite{16}. Re-writing the constants of motion given in \cite{14} the contracted Lie algebra is obtained to be
\begin{equation}
[x_0, p]= i \hbar
\end{equation}
\begin{equation}
[q_0, L ] =-i \hbar p_0
\end{equation}
\begin{equation}
[p_0, L] = i \hbar q_0
\end {equation}
The dimensionless variables are $x_0 \rightarrow q_0, ~ p \rightarrow p_0$; the variable $y_0$ in the dimension of momentum could be written as $p=m \omega_c y_0$ 
\begin{equation}
x_0 =\frac{x}{2} - \frac{p_y}{m \omega_c}
\end{equation}
\begin{equation}
y_0 =\frac{y}{2} + \frac{p_x}{m \omega_c}
\end{equation}
The Lie algebra (24) -(26) defines the dynamical symmetry group of the Landau problem; it is that of the central extension of the Euclidean group $\bar{E}(2)$.

Introducing another set of canonically conjugate variables
\begin{equation}
X=\frac{x}{2} + \frac{p_y}{m \omega_c}
\end{equation}
\begin{equation}
P = p_x - \frac{m\omega_c y}{2}
\end{equation}
The commutator is
\begin{equation}
[X, P] =i \hbar
\end{equation}
and the Hamiltonian (9) in terms of $(X, P)$ reads
\begin{equation}
H_L = \frac{P^2}{2m} + \frac{1}{2} m \omega_c^2 X^2
\end{equation}
The Heisenberg algebra (31) completed to "strictly quadratic" algebra \cite{21} is obtained defining
\begin{equation}
I_1 = \frac{1}{4} (XP+PX)
\end{equation}
\begin{equation}
I_2 = \frac{1}{4} (X^2 - P^2)
\end{equation}
\begin{equation}
I_3 =-\frac{1}{4} (X^2 + P^2)
\end{equation}
we get the following commutators in the units $\hbar=1$
\begin{equation}
[I_1, I_2] =i I_3, ~ [I_2, I_3] =-i I_1, ~ [I_3, I_1] = -i I_2
\end{equation}
The algebra (36) is the Lie algebra of the group $SL(2, R)\sim Sp(2, R) \sim SO(2,1)$. The group SL(2, R) is the spectrum generating group; the generators $I_j, j=1,2,3$ do not commute with the Hamiltonian (32). The new result using group contraction of SU(2) is very insightful: the contraction process in two different ways leads to Heisenberg algebra whose completion determines the spectrum generating group Sp(2, R) isomorphic to SL(2, R), and to the algebra of the group E(2) whose central extension leads to the group $\bar{E}(2)$ that determines the dynamical symmetry of the Landau problem. 

The symmetry groups are independent of the chosen gauge, i. e. symmetric or Landau, however the gauge symmetry plays important role in this gauge-independence. The idea of Wakamatsu and Hayashi \cite{8} to seek physical interpretation from the calculation of matrix elements is interesting. To put this work in the perspective of our approach based on symmetry groups we recall the important results of \cite{24} and remarks on the role of gauge transformations in \cite{14}. The classical equations of cyclotron motion lead directly to the constants of motion $(x_0, y_0)$  and 1D HO energy eigenvalues and eigenfunctions are derived for the Landau problem in \cite{24}. The conserved momenta $(p_x^{cons},~p_y^{cons})$ and orbital angular momentum in \cite{8} are exactly the same as given in \cite{24}. Dulock and McIntosh \cite{14} point out that the gauge transformation is intimately related with the space translation in 2D plane.  Thus, the role of the physical picture of the Landau motion in 2D plane in which the center of the orbit and the location of the orbit are crucial is common to all the considerations given in \cite{8,14,16,24}. The physical import of pure gauge potential proposed in \cite{25,26} presents a departure from these interpretations.

Let us consider the pure gauge field
\begin{equation}
{\bf A}^{pure} = {\bf \nabla} \chi ;~ \chi = \frac{1}{2} B x y
\end{equation}
The pure gauge field has nonvanishing momentum as well as orbital angular momentum ${\bf r} \times {\bf A}$
\begin{equation}
{\bf p}^{pure} = \frac{e B}{2c} (y \hat{x} +x \hat{y})
\end{equation}
\begin{equation}
L^{pure}_z = \frac{e B}{2 c} (x^2 - y^2)
\end{equation}
The uniform magnetic field is unaltered by the gauge transformation ${\bf A} \rightarrow {\bf A} \pm {\bf A}^{pure}$, however the contribution of pure gauge field to momentum and orbital angular momentum is nonzero and differs depending on the choice of the gauge. Note that one can also calculate  ${\bf \nabla }(r^2 \sin \phi \cos \phi)$ in polar coordinates for pure gauge field in Eqs. (38-39).  How do we understand the claimed redundancy of the gauge symmetry \cite{8}? 

The answer depends on the physical reality of the electromagetic potential. Motion of an electron in a constant magnetic field along z-axis spread uniformly over the 2D plane does not need vector potential for its description; the vector potential itself would be redundant. The divergence equation  ${\bf \nabla}.{\bf B} =0$ implies that one can introduce ${\bf A}$ to define ${\bf B} = {\bf \nabla} \times {\bf A}$ and then one has the idea of gauge symmetry. If the magnetic field is a constant pseudo-vector the divergence equation is an identity and the preceding prescription to introduce ${\bf A}$ is vacuous. However, proceeding in the reverse order postulating physical reality of the vector potential and defining ${\bf B} = {\bf \nabla} \times {\bf A}$ the construction of a constant ${\bf B}$ becomes a logically valid issue as is done in the Landau problem. Physical implications would follow if the translational symmetry is related with the gauge symmetry \cite{14,16}. In quantum theory it would seem that though the vector potential is present in the Schroedinger equation the gauge invariance in the extended form of phase invariance of the wavefunction makes the physical reality of the vector potential debatable \cite{27, 28}. In the case of the Aharonov-Bohm effect \cite{27} instead of a 2D plane it is possible to seek the physical meaning of ${\bf A}$ using the punctured 2D plane, $R^2 -{0}$, and use integral form for defining ${\bf B}$ instead of the divergence law, i. e. the modified Stokes theorem \cite{28}. In contrast to the conclusion of \cite{8} we find that the gauge symmetry has important physical role in the Landau problem: the emergence of the Euclidean group and the space translation symmetry linked with the gauge transformation.

\section{\bf SO(3,2) symmetry group and generalized Landau problem}

Manifest Lorentz covariance and the question of longitudinal and time-like photons in the canonical quantization using the condition (5) equivalent to $p^\mu A_\mu =0$ has been of concern to the author in connection with the decomposition of angular momentum into spin and orbital parts \cite{2}. Recent discussions \cite{12,13} throw light on this issue. We may, however, seek more insights using wider symmetry group for the extended structure of proton. Dirac's remarkable representaion of the group SO(3,2) \cite{15} has the important property that the group contraction resulting into the Poincare group has irreducible unitary representations that have positive-definite or negative-definite energies. In an interesting paper \cite{29} quantum relativistic oscillator for extended particles (hadrons) is treated using the group SO(3,2), and the nonrelativistic limit of the internal motion is obtained. Dirac himself used a two-oscillator system for the Majorana representaion of the positive energy relativistic equation \cite{30, 31} based on SO(3,2) group. Is it possible to generalize the Landau problem that has SO(3,2) symmetry group? In this section, we address this question. 

Let us summarize rudiments of Dirac's remarkable representation  first. The group of rotations that leaves the quadratic form
\begin{equation}
g_{ab} x^a x^b = x_1^2 +x_2^2 + x_3^2 - x_4^2 - x_5^2
\end{equation}
invariant defines the 3+2 de Sitter group SO(3,2). Here the indices $a, b = 1,2,3,4,5$ and $\mu, \nu =1,2,3,4$. The antisymmetric generators  $m_{ab}=-m_{ba}$ satisfy the commutation relations
\begin{equation}
[m_{ab}, m_{cd}] =0
\end{equation}
if $a,b,c,d$ are all different, and
\begin{equation}
[m_{ab}, m_{ac}] =m_{bc} ~ ~ a=4,5
\end{equation}
\begin{equation}
[m_{ab}, m_{ac}] = -m_{bc} ~~ a=1,2,3
\end{equation}
where $a,b,c$ are all different. Note that $i m_{ab}$ for all $a,b$ have real eigenvalues for unitary representations.  Introducing a set of real variables $(q_1, p_1, q_2, p_2)$ to get the expressions for all $i m^\prime_{ab}$, it is found that they satisfy the commutators $[q_i, p_j] =i \delta_{ij}, ~i,j =1,2$ similar to those of canonically conjugate variables in quantum mechanics.  Four of the ten generators are cyclic
\begin{equation}
i m^\prime_{12} = \frac{1}{2} (q_1 p_2 - q_2 p_1)
\end{equation}
\begin{equation}
i m^\prime_{23} = \frac{1}{4} (q_1^2 + p_1^2 - q_2^2  -p_2^2)
\end{equation}
\begin{equation}
i m^\prime_{31} = -\frac{1}{2}(q_1 q_2 + p_1 p_2)
\end{equation}
and $i m^\prime_{45}$ is given by 
\begin{equation}
i m^\prime_{45} = \frac{1}{4} (q_1^2 + p_1^2 +q_2^2+p_2^2)
\end{equation}
It is interesting to note that $i m^\prime_{45}$ resembles the half of the sum of the energy of two 1D oscillators; the angular momentum operator $i m^\prime_{12}$ has half-odd eigenvalues for odd wavefunctions, and the energy eigenvalues for $i m^\prime_{45}$ have integral positive values. The set of the operators $(i m^\prime_{12}, im^\prime_{23}, i m^\prime_{13})$ satisfies the sub-algebra isomorphic to the SU(2) algebra whereas the set of the generators $(i m^\prime_{45}, i m^\prime_{43}. i m^\prime_{35})$ satisfies a sub-algebra isomorphic to that of the concompact group SO(2,1) equivalent to Sp(2, R).  

We have explained in \cite{16} that the Dirac two-oscillator system cannot be a 2D IHO or two independent 1D HOs. In an important work Baskal et al \cite{17} point out that though one can construct 16 quadratics for a two-oscillator system only 10 are needed for the Lie algebra of the group SO(3,2). In the last part of his paper \cite{15} Dirac also proved the equivalence of the group SO(3,2) with the symplectic group Sp(4, R). The modest aim that we have set here is to generalize the Landau problem such that the use of the group SO(3,2) becomes applicable. It is a nontrivial task; we seek guidance from two directions. Jauch and Hill \cite{19} gave examples of the dynamical symmetry groups SO(3) and SO(2,1) for the 2D Kepler-Coulomb problem and SU(2) for 2D IHO; in the former case because of the possibility of a repulsive force and the continuum of scattering energy states the Lorentz group SO(2,1) occurs as a symmetry group. For a full correspondence between two systems it would be natural to consider 2D IHO with a repulsive potential, i. e. the usual correspondence \cite{25} has to be extended. From another interesting angle Kibler \cite{18} notes that for N-dimensional IHO, $N \geq 1$,  the dynamical non-invariance group is the symplectic group Sp(2N, R). Its sub-group SU(N) is a maximal compact dynamical symmetry group. For N=2 to get the group Sp(4,R) we need 2D IHO with an attractive as well as a repulsive potential. Now, Dirac's proof for the equivalence of symplectic and de Sitter groups implies $Sp(4,R) \rightarrow SO(3,2)$.

To understand the importance of a repulsive potential let us consider a 1D system described by the following Hamiltonian
\begin{equation}
H_{1r}= \frac{1}{2m} p^2 - \frac{m \omega^2}{2} x^2
\end{equation}
Here "r" in the subscript of the Hamiltonian signifies repulsive potential that represents the scattering from a parabolic potential. One could get symplectic group Sp(2,R) using the transformations $(x \rightarrow x \cosh \beta + p \sinh \beta ;~ p \rightarrow x \sinh \beta  + p \cosh \beta)$ where $\beta = \sqrt{ \frac{m \omega}{\hbar}}$. Interestingly the squeezing and dilation generators in Sp(2,R) could be viewed in terms of the Lorentz transformations in (2+1)- D space-time. 

It becomes clear that the argument used in \cite{17} to interpret four of the ten generators considering the 4D phase space volume preserving generators could be related with the repulsive potential discussed in \cite{18}. Since $Sp(2,R) \sim SO(2,1)$ and the generators $(i m^\prime_{45}, i m^\prime_{43}. i m^\prime_{35})$ of SO(3,2) satisfy a sub-algebra isomorphic to that of the noncompact group SO(2,1) a 2D oscillator system with both attractive and repulsive potentials will have SO(3,2) symmetry. We are interested, however, in the Landau problem: adopting the Landau-Zeeman paradigm, and keeping intact the Landau part, a generalized system for SO(3,2) symmetry group could be envisaged introducing a modified Hamiltonian
\begin{equation}
H_Z^R = H_L -\frac{1}{2} m \omega^2 (x^2 + y^2)
\end{equation}
Evidently the full symmetry group will be $Sp(4,R) \sim SO(3,2)$. We may construct a single generalized Hamiltonian with attractive and repulsive potentials having the frequency/energy $\omega$ and $\Omega$ respectively
\begin{equation}
H_Z^G = H_L +\frac{1}{2} m \omega^2 (x^2 + y^2)-\frac{1}{2} m \Omega^2 (x^2 + y^2)
\end{equation}
and study this system for two cases $\omega^2 >\Omega^2$ and  $\omega^2 <\Omega^2$.

Let us consider the modified Hamiltonian $H_Z^R$ and calculate the Poisson brackets
\begin{equation}
[H_Z^R,~ q_i]= a_{ij} q_j
\end{equation}
where $q_i =x,p_x,y,p_y; ~i=1,2,3,4$. We get the matrix representation of the Hamiltonian
\begin{equation} 
H_Z^R =\begin{bmatrix} 0 & \omega_L & m(\omega_L^2 -\omega^2) & 0 \\ -\omega_L & 0 & 0 & m(\omega_L^2 -\omega^2) \\ -\frac{1}{m} & 0 & 0 & \omega_L \\ 0 & -\frac{1}{m} &  -\omega_L & 0 \end{bmatrix}
\end{equation}

The calculated eigenvalues of the Hamiltonian in the matrix representation are
\begin{equation}
\pm i [(\omega_L^2 -\omega^2)^\frac{1}{2} +\omega_L]; ~\pm i [(\omega_L^2 -\omega^2)^\frac{1}{2} -\omega_L]
\end{equation}
We find three important special cases from the set of four eigenvalues (53). Case I: If the Larmor frequency is greater than $\omega$ then the modified Landau-Zeeman problem is equivalent to the Landau-Zeeman problem discussed in \cite{14} and the generators defined by (18)-(20) determine the symmetry group to be SU(2). Case II: If $\omega_L =\omega$ the four eigenvalues degenerate to two
\begin{equation}
E_{osc} =\pm i \omega_L
\end{equation}
Here the eigenvalues and eigenfunctions correspond to 2D IHO, and the symmetry group is SU(2). Case III: Taking the limit $\omega_L \rightarrow 0$ the degenerate eigenvalues are
\begin{equation}
E_{osc}^R =\pm \omega
\end{equation}
In contrast to the  2D IHO obtained in section 3 for the Landau-Zeeman Hamiltonian in this limit, here we get 2D isotropic oscillator with a repulsive potential and the symmetry group $Sp(2,R) \sim SO(2,1)$.

Thus the relevant symmetry group for the typical modified Landau-Zeeman Hamiltonian  $H_Z^R$  proposed here considering the Case II and Case III is Sp(4,R). In view of Dirac's proof that SO(3,2) is equivalent to Sp(4,R) \cite{15} the remarkable representation of SO(3,2) becomes natural for this problem.

\section{\bf Gauge invariance and proton spin problem: a new perspective}

It seems the symmetry of the Landau problem discussed here may turn out to be important for the proton spin problem. First, we make few remarks on the recent approach on the Landau problem \cite{8}. The derivation of the set of four eigenfunctions  $\Psi^{(S)}_{n,m}; ~ \Psi^{(S)}_{n,k_x}; ~\Psi^{(L_1)}_{n,m}; ~\Psi^{(L_1)}_{n,k_x}$ given in \cite{8} is not difficult, however, the idea to calculate various matrix elements to gain physical insights is interesting and technically involved. In the case of the Landau-Zeeman Hamiltonian (13) it would be interesting to follow this approach and calculate matrix elements. Somewhat more difficult, but having the potential to gain new insights  would be  to calculate matrix elements for the generalized Landau-Zeeman Hamiltonian (49). It may be pointed out in this connection that in analogy to 1D HO, the solution of the time-independent Schroedinger equation $H_{1r} =E \Psi$ has been obtained by Shimbori and Kobayashi \cite{32} in terms of complex eigenvalues and the eiegnfunctions $w_n(x) \propto e^{i x^2/2} H_n(x)$ where
\begin{equation}
\frac{d^2 H_n}{d x^2} +2 i x \frac{d H_n}{d x} -2 i n H_n =0
\end{equation}

 In quantum field theory it is known that for the massless particles the space-time symmetries can be described in terms of Wigner's little group that is isomorphic to E(2). Kim and Wigner \cite{33} introduce cylindrical group as a contraction group of the rotation group O(3) following \cite{20}, and also show that the gauge transformation for photon (or massless particles) could be interpreted in terms of Lorentz-boosted rotations. Linking translation symmetry of E(2) with the gauge transformations for photon field is a remarkable result \cite{33} that makes the implication of the Landau problem for QED and QCD more relevant. Little group arises from the group of Lorentz transformations $L^\mu_\nu$ applied to a light-like 4-vector $K^\mu$
\begin{equation} 
L^\mu_\nu K^\nu = K^\mu
\end{equation}
For a single photon propagating along z-axis we have $K^1 =K^2 =0 ; ~K^3 =k, K^0 =k c$. The Lorentz transformation matrix in this case can be factored into a roatation matrix and translations in E(2): $R(\alpha) D(u,v)$. Explicitly, for the rotation parameter $\alpha$ and translation parameters $(u,v)$ for E(2) we have
\begin{equation}
L^\mu_\nu = \begin{bmatrix} \cos \alpha & \sin \alpha & 0 & 0 \\ -\sin \alpha & \cos \alpha & 0 & 0 \\ 0 & 0 & 1 & 0 \\ 0 & 0 & 0 & 1 \end{bmatrix} \begin{bmatrix} 1 & 0 & -u & u \\ 0 & 1 & -v & v \\ u & v & 1 -(u^2 +v^2)/2 & (u^2 +v^2)/2 \\  u & v & -(u^2 +v^2)/2 & 1 +(u^2 +v^2)/2 \end{bmatrix}
\end{equation}
Imposing the Lorentz condition for the vector potential $\partial^\mu A_\mu = p^\mu A_\mu =0$ the $4 \times 4$ representation of the little group is reduced to block diagonal form in which one gets the generators of rotation and E(2).

Above result on the little group has, in fact, direct relation with quantum field theory as shown by Weinberg \cite{34}. The invariance of S-matrix for the emission of a massless particle (e. g. photon) under infinitesimal Lorentz transformations is related with the mass-shell gauge invariance making use of the matrix (58) and re-gauging the polarization vector \cite{34}. Conceptually, we expect new perspective on the calculation of the matrix elements for the gauge-fixing covariant operators in \cite{11, 12, 13}. In the book \cite{23} Weinberg follows this approach in Chapter 5. However, though speculative, the idea of Kim and Wigner \cite{33} motivated by \cite{34} could be more fruitful in the present context. Since gluon is also a massless field we suggest to re-visit QCD in connection with proton spin problem in this perspective. Of course, Weinberg's arguments \cite{34} are limited to photon and graviton, and generalization to nonabelien gluon gauge field is not obvious.

Now, gauge symmetry from the point of view of a general conceptual framework that wider symmetry groups provide is discussed based on Dirac's two oscillator system. The two-oscillator system in \cite{30,31} is described by six (anti-symmetric) generators $s_{\mu\nu}$ corresponding to spin operators of the Lorentz group, and four additional generators $s_{\mu 5}$ to label the irreducible representations thus having the equivalence with 10 generators of the remarkable representation of SO(3,2) \cite{15}. Speculative picture of a particle in de Sitter space-time \cite{31} in a special case with zero momentum has circular motion, and the motion for different Hamiltonians is described by plane sections of a spherical shell. We elaborate on this model for clarity emphasizing aspects related with the two-oscillator system and angular momentum.

Considering the positive energy relativistic wave equation, i. e. the Majorana equation \cite{30}, the generators of the infinitesimal rotations give the spin angular momentum operators $s_{\mu\nu}$ which could be related with the canonical variables of two oscillators in the remarkable representation of the group SO(3,2) \cite{15}. Mathematically, additional four quantities $s_{\mu 5} =-s_{5 \mu}$ are introduced to get 10 generators $m_{ab} =-m_{ba}$ for the group SO(3,2) defined here in section 5. The Majorana equation has the Hamiltonian
\begin{equation}
H_5 = 1 -2 s_{\mu 5} p^\mu
\end{equation}
One of the important results is that the usual prescription $p_\mu \rightarrow p_\mu +\frac{i e}{c} A_\mu$ for introducing the interaction is consistent only for pure gauge field $F_{\mu\nu} =0$. In his second paper, a particle model is developed solving the Heisenberg equations of motion \cite{31}.

For $H_5$ and time variable $\tau_5$ the solution is given by
\begin{equation}
s_{r5} =b_r \cos 2\tau_5 +b_r^\prime \sin 2\tau_5
\end{equation}
\begin{equation}
s_{r0} =-b_r \sin 2\tau_5 +b_r^\prime \cos 2\tau_5
\end{equation}
The constants $b_r,~ b_r^\prime$ are the initial values of $s_{r5},~ s_{r0}$ respectively. The Heisenberg equations of motion for other Hamiltonians $H_0, H_1, H_2, H_3$ are also solved; for $H_0$ there is no change in the dynamical variables. 

The remaining three Hamiltonians $ H_1, H_2, H_3$ describe circular motions. A constant time variable $x_0$ and the coordinates $x_r, ~r=1,2,3$ describe circular motion. Dirac defines new variables
\begin{equation}
y_r =x_r +s_{r0}
\end{equation}
In this picture $s_{r0}$ oscillates into $s_{r5}$ with time, the point $x_r$ roams around all over the shell centered at $y_r$ and radius vector $-s_{r0}$, and this motion is like a gauge transformation. For a zero momentum particle the internal motion is described by a pulsating spherical shell with the coordinates of the centre $y_r$. The centre is fixed
\begin{equation}
\frac{d y_r}{d \tau_5} =0
\end{equation}
the motion of a point $x_r$ is a gauge transformation.

For an arbitrary non-zero momentum the solution for $y_\mu$ of the equation
\begin{equation}
\frac{d y_\mu}{d \tau_5} =p_\mu
\end{equation}
is $y_\mu =p_\mu \tau_5 +c_\mu$, $c_\mu$ is constant. Now, the $y_\mu$ describe a world-line in 4D space-time, and the point $x_\mu$ roams over the surface of a tube. However, the spin is momentum dependent.

The momentum-dependent spin is unphysical and Dirac argues that (i) to define spin unambiguously one needs a definition of orbital angular momentum that depends on some coordinates, (ii) for zero momentum the choice of coordinates is immaterial, and (iii) using gauge transformation one can define new coordinates such that unambiguous spin could be defined. In this particular case, orbital angular momentum using $x_r$ 
\begin{equation}
m_{rs} = x_r p_s -  x_s p_r
\end{equation}
would give unphysical spin. However, using $y_r$ in (65) zero spin for all values of momentum is obtained by Dirac. 

It is known that the new relativistic wave equation discussed in \cite{30,31} does not describe physical electron. However, the irreducible representations of the group SO(3,2) have been of interest in hadron models. Bohm et al \cite{29} discuss a relativistic particle with internal structure using the group contraction of the group SO(3,2), and note two sub-group chains for irreducible representations of SO(3,2) group.
Note that the vector $\Gamma^\mu$ in \cite{29} corresponds to Dirac's $m_{\mu 5}\equiv s_{\mu 5}$. The quadratic Casimir invariant $C_2$ and the fourth-order invariant $C_4$ have the values $-\frac{5}{4}, ~0$ respectively for the Dirac representation \cite{15}.  Authors \cite{29} derive an important result for a special class of irreducible representations of SO(3,2): the orbital angular momentum of the oscillator itself defines spin of the states. 

Two important physical results could be obtained from the group theoretic considerations based on the central extension of the Euclidean group for the Landau problem, $Sp(4,R) \sim SO(3,2)$ for the generalized Landau-Zeeman problem, the Euclidean little group for photon, and SO(3,2) for Dirac's two oscillator system. First, the translational symmetry is intimately linked with the gauge symmetry. Second, pure gauge field arises naturally in all cases, and the choice of a coordinate system has to be such that there exists a fixed point (for example, $y_r$) or a definite world-line, for example, defined by $y_\mu$. This amounts to some kind of field singularity and/or nontrivial topology. In the proton spin decomposition controversy we obtain following insights.

{\bf I: Decomposition of the gauge potential}

One of the most debated issues concerns the proposed splitting of the gauge potential into pure gauge and physical components \cite{1}. With a remarkable clarity \cite{3,7,8} it has been argued that this splitting amounts to longitudinal and transverse components of the vector field ${\bf A}$, and it is unique:  the Coulomb gauge and the fixed Lorentz frame. In fact, a more direct approach is to use the Coulomb gauge straightway \cite{23}. If manifest Lorentz covariance is insisted upon there arise unresolved issues. In the light of the present results we re-examine the nature of the pure gauge vector potential (37). It is easy to verify that
\begin{equation}
{\bf \nabla}.{\bf A}^{pure}={\bf \nabla} \times {\bf A}^{pure}=0
\end{equation}
It is a harmonic vector field, nonsingular in the whole domain $R^3$, and there is no nontrivial topology. Therefore, the arguments based on the Landau problem \cite{7,8} cannot settle the issue. The Euclidean group E(2) arises because of the uniform spread of ${\bf B}$ in the 2D plane for the Landau problem, and there does not exist a fixed centre of the Landau orbit. For the SO(3,2) symmetry group we have pointed out that Dirac used a fixed point to define spin. Instead of a uniform magnetic field Landau problem we suggest confined magnetic flux for the Aharonov-Bohm effect that requires q-fold covering group of E(2) \cite{28} to address this question. Recalling the Hodge decomposition of a vector field for de Rham periods \cite{10,35}
\begin{equation}
{\bf A} = {\bf A}_{exact} +{\bf A}_{closed} + {\bf A}_{harmonic}
\end{equation}
where ${\bf A}_{exact}$ has zero curl but non-zero divergence, ${\bf A}_{closed}$ has zero divergence but non-zero curl, and ${\bf A}_{harmonic}$ has both divergence and curl zero. We propose the harmonic vector field to be a pure gauge field; illustrative example is as follows
\begin{equation}
{\bf A}^{pure} = {\bf A}_{harmonic} = \frac{y}{x^2+y^2} \hat{x} -  \frac{x}{x^2+y^2} \hat{y}+0 \hat{z}
\end{equation}
Nontrivial topology of the punctured plane $R^2 -\{0\}$ having fixed origin determines the pure gauge potential uniquely. Though application of this idea for the gluon field in QCD is not easy, the approach discussed in \cite{10} could be re-visited in this perspective.

{\bf II: Angular momentum decomposition}

Recently following decomposition for the canonical angular momentum in QED has been suggested \cite{12}
\begin{equation}
{\bf J} = {\bf J}_{spin} + {\bf J}_{orb} + {\bf J}_{gf}
\end{equation}
Canonical versus mechanical, and symmetric versus nonsymmetric angular momentum in connection with the physical angular momentum in quantum optics, QED and QCD have been extensively discussed in the literature, see reviews \cite{4,5} for references. The last term ${\bf J}_{gf}$ arising from the gauge fixing is a new addition \cite{12}. The argument is to obtain correct matrix elements for physical operators. To understand the nature of the angular momentum decomposition we recall the topological arguments \cite{10} leading to following decomposition
\begin{equation}
{\bf J} = {\bf J}_{spin} + {\bf J}_{orb} + {\bf J}_{top}
\end{equation}
Topological part ${\bf J}_{top}$ acquires significance in the context of SO(3,2) group where spin and orbital parts may loose their distinction for certain irreducible representations \cite{29,30,31}. Since the time component of the vector potential $A_0$ appears in the integrand of 
$ {\bf J}_{gf}$ it is not obvious if the gauge-fixing angular momentum could be related with topological one. However, the role of pure divergence term relating canonical and symmetric angular momentum may have topological implications. Could there be some link between Berry phase, angular momentum holonomy conjecture \cite{36}, and gauge-fixing angular momentum \cite{12}? We leave this question for future work.

\section{\bf Conclusion}

The extended bound structure of proton and the scattering experiments have led to the unresolved proton spin problem. An important and controversial question relates with the role of gauge symmetry in the definition of physical spin and orbital angular momentum of the constituents of proton, i. e. quarks and gluons. Role of gauge symmetry in the Landau problem has been explored in the literature to gain insights on the proton spin problem. In this letter we have established the importance of gauge symmetry for the 2D space translation symmetry in the Landau problem. A generalized Landau-Zeeman system having SO(3,2) symmetry group is proposed, and the importance of gauge transformation for the definition of spin and orbital angular momentum based on Dirac two-oscillator system is discussed. It is argued that to settle the controversy on the angular momentum decompositions one needs group theoretic and topological considerations.

\end{document}